\documentclass[10pt,aps,prl,twocolumn,superscriptaddress]{revtex4}

\usepackage[utf8]{inputenc}
\usepackage{hyperref}
\usepackage{graphicx}
\usepackage{amsmath,amssymb}
\usepackage{color}
\usepackage{subfigure}
\usepackage{url}
\usepackage{flushend}
\graphicspath{{Images/}}

\begin{document}

\title{
Phase-resolved heterodyne holographic vibrometry with a strobe local oscillator
}

\author{Nicolas Verrier}
\author{Michel Gross}
\affiliation{Laboratoire Charles Coulomb, CNRS UMR 5221, Universit\'e Montpellier II. 34095 Montpellier, France}
\author{Michael Atlan}
\affiliation{
Institut Langevin. Fondation Pierre-Gilles de Gennes. Centre National de la Recherche Scientifique (CNRS) UMR 7587, Institut National de la Sant\'e et de la Recherche M\'edicale (INSERM) U 979, Universit\'e Pierre et Marie Curie (UPMC), Universit\'e Paris 7. \'Ecole Sup\'erieure de Physique et de Chimie Industrielles - 1 rue Jussieu. 75005 Paris. France
}

\date{\today}

\begin{abstract}
We report a demonstration of phase-resolved vibrometry, in which out-of-plane sinusoidal motion is assessed by heterodyne holography. In heterodyne holography, the beam in the reference channel is an optical local oscillator (LO). It is frequency-shifted with respect to the illumination beam to enable frequency conversion within the sensor bandwidth. The proposed scheme introduces a strobe LO, where the reference beam is frequency-shifted and modulated in amplitude, to alleviate the issue of phase retrieval. The strobe LO is both tuned around the first optical modulation side band at the vibration frequency, and modulated in amplitude to freeze selected mechanical vibration states sequentially. The phase map of the vibration can then be derived from the demodulation of successive vibration states.
\end{abstract}
\maketitle

Time-averaged holography is an efficient detection method for mapping sinusoidal out-of-plane vibration amplitudes. Homodyne \cite{Powell1965, Levitt1976, PicartLeval2003, Pedrini2006} and heterodyne \cite{Aleksoff1971, JoudLaloe2009} modalities were investigated. However, achieving a robust mechanical phase mapping remains an important issue. In holography, the local phase retardation of the scattered optical field impinging onto each pixel of a sensor array is used for image reconstruction. Nevertheless, it is not exploited straightforwardly for the determination of the local mechanical phase of an object in sinusoidal motion unless if a high framerate sensor is used \cite{Pedrini2006}. A stroboscopic approach can alleviate this issue. Several holographic and interferometric approaches have been proposed and validated for mechanical phase retrieval from sequences of intensity images under phase-stepped stroboscopic illumination \cite{Lokberg1976, Petitgrand2001, LevalPicart2005, Hernandez2011}.\\

In this letter, we report the use of a stroboscopic scheme in conjunction with frequency-shifted modulation sideband detection to map mechanical phase shifts with respect to the excitation signal at a given vibration frequency. Detuning the strobe frequency with respect to the mechanical excitation signal by a fraction of the sampling frequency can be used to freeze $m$ linearly-spaced vibration states over a vibration period. Each of these states can then be measured, from sequentially-acquired frames. Furthermore, detuning the carrier frequency of the reference beam with respect to the object beam will enable phase-shifting holography with optimal accuracy \cite{AtlanGross2007}. Appropriate LO carrier frequency will provoke $n$-phase modulation of the optical signal from each of the $m$ vibration states. The purpose of this scheme is to recover a phase map of the vibration from the demodulation of $m$ intensity holographic images, each of which will be derived from the demodulation of $n$ raw interferograms. Hence a phase map measurement will require $m \times n$ sequentially recorded frames.\\
 
\begin{figure}[t]
\centering
\includegraphics[width = 8 cm]{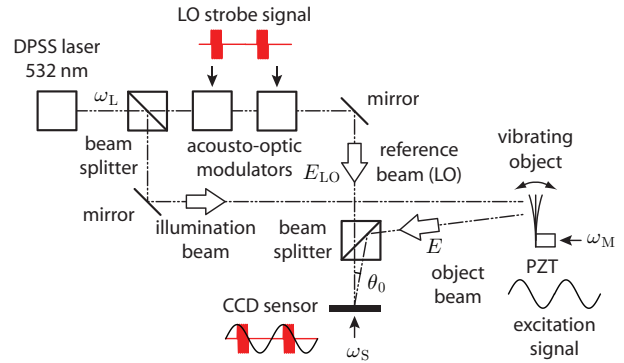}
\caption{Sketch of the optical setup.}
\label{fig_setup}
\end{figure}

The optical acquisition set-up, sketched in Fig. \ref{fig_setup}, consists of an off-axis heterodyne Mach-Zehnder interferometer allowing detection of modulation sidebands of the optical radiation scattered by a vibrating object. This holographic interferometer is used for the detection of an object field $E$ in reflective geometry, beating against a LO field $E_{\rm LO}$. The optical radiation is provided by a 100 mW, single-mode, doubled Nd:YAG laser (Oxxius SLIM 532) at wavelength $\lambda = 532$ nm, and optical frequency $\omega_{\rm L} / (2 \pi) = 5.6 \times 10^{14} \, \rm Hz$. The LO field is shaped by two acousto-optic modulators (AA-electronics, MT80-A1.5-VIS). It has the form $E_{\rm LO} = {\cal E}_{\rm LO} {e} ^{i (\omega_{\rm L} + \Delta \omega) t} H_{\rm AM} (t)$, where $\cal E_{\rm LO}$ is its complex amplitude. The frequency shift of the carrier is $\Delta \omega$, and $H_{\rm AM} (t)$ is the amplitude modulation (AM) function of the LO pulses : a square signal with a cyclic ratio of 25\%; its angular frequency is $\omega_{\rm P}$. The studied objects are attached to a piezo-electric actuator (Thorlabs AE0505D08F), vibrating sinusoidally, at angular frequency $\omega_{\rm M}$. The out-of-plane vibration $z \sin(\omega_{\rm M} t + \psi)$, of amplitude $z$ and phase retardation $\psi$, provokes a modulation of the optical path length of the object field $E = {\cal E} {e} ^{i \omega_{\rm L} t + i \phi(t)}$, where ${\cal E}$ is its complex amplitude. The phase modulation of the backscattered light is $\phi(t) = \phi_{0} \sin(\omega_{\rm M} t + \psi)$, where $\phi_{0} = 4 \pi z / \lambda$ is the modulation depth, which generates radiofrequency optical modulation side bands at harmonics of $\omega_{\rm M}$. For each mechanical phase map retrieval at a given excitation frequency, a set of $m\times n$ optical interferograms ${\cal I}_k$ ($k=1,\ldots, mn$) is recorded by an Andor EMCCD sensor array of $1002 \times 1004$ pixels. Throughout the experiments, raw frames are sampled at a frame rate of $\omega_{\rm S} / (2 \pi) = 24 \, \rm Hz$. Holographic image rendering from the recorded interferograms ${\cal I}_k$ at times $2 \pi k / \omega_{\rm S}$ involves a scalar diffraction calculation. It is performed with a numerical Fresnel transform ${\cal F}$, a special case of linear canonical transform, whose digital computing is studied in detail in \cite{HennellySheridan2005}. Its implementation in off-axis holographic conditions \cite{SamsonVerpillat2011} yields holograms $I_k$ reconstructed in the object plane.\\

\begin{figure}[t]
\centering
\includegraphics[width = 7.0 cm]{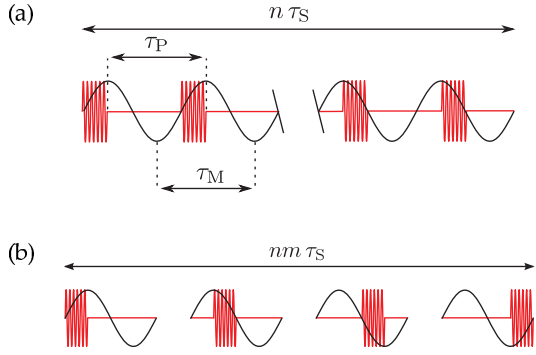}
\caption{(Color on-line) Chronogram of signal acquisition over $n$ and $m \times n$ successive frames, (a) and (b), respectively. Black: mechanical excitation signal, red: strobe LO signal.}
\label{fig_chronogram}
\end{figure}

Signal modulation consists in choosing appropriate LO pulse and carrier frequencies, $\omega _{\rm P}$ and $\Delta \omega$, respectively. We want the LO pulse pattern of period $\tau _{\rm P} = 2 \pi / \omega _{\rm P}$, shown in fig. \ref{fig_chronogram}(a) and fig. \ref{fig_chronogram}(b), to be shifted by one $m$-th of the mechanical signal pattern (of period $\tau _{\rm M} = 2 \pi / \omega _{\rm M}$) after $n$ consecutive images. This implies
$\tau _{\rm P} - \tau _{\rm M} = \tau_{\rm{M}}^2 / (n m  \tau_{\rm{S}})$, where $\tau_{\rm{S}}=2\pi/\omega_{\rm{S}}$ is the time elapsed from-frame-to-frame recording. For pulse frequencies close to the excitation frequency $\tau_{\rm{P}} - \tau_{\rm{M}} \ll \tau_{\rm{M}}$, to provoke the desired drift between the optical and the mechanical signal pattern (Fig.\ref{fig_chronogram}(a)), the pulse frequency is set to 
\begin{equation}\label{eq_StrobeFreq}
\omega_{\rm P} = \omega_{\rm M} + \omega_{\rm{S}} / (n m)
\end{equation}
Furthermore, the LO carrier is detuned by $\Delta \omega = \omega_{M} + \omega_{\rm S} / n$ to provoke a beat of the first optical modulation side band component in the recorded interferogram at $\omega_{\rm S} / n$. Following these guidelines, signal demodulation is made in two steps. First, a set of $m \times n$ consecutive raw intensity images $I_k$ ($k=1,\ldots, mn$) is acquired (Fig. \ref{fig_chronogram}(b)), from which $m$ states $M_p$ ($p=1,\ldots, m$) within the mechanical excitation period are derived 
\begin{equation}\label{eq_M}
M_p=\sum_{k=1}^n I_{k+(p-1)n} \exp\left({-\frac{2i\pi k}{n}}\right).
\end{equation}
The phase information of intermediate complex-valued holograms $M_p$ is discarded. Intermediate images $\left|M_p\right|^2 \propto z_0 \sin ( 2 p \pi \omega_{\rm M} / \omega_{\rm S} + \psi)$ are used as new data, from which a local vibration map $Q$ is demodulated
\begin{equation}\label{eq_Q}
Q=\sum_{p=1}^m\left|M_p\right|^2\exp\left({-\frac{2i\pi p}{m}}\right).
\end{equation}
For small vibrations ($z \ll \lambda$), the magnitude of the quantity derived in eq. \ref{eq_Q} is proportional to the local out-of-plane vibration amplitude $\left|Q\right| \propto z$. Its argument yields the local phase retardation with respect to the excitation signal $\psi \equiv {\rm arg}\left(Q\right) \left[ 2 \pi \right]$.\\ 

\begin{figure}[t]
\centering
\includegraphics[width = 8 cm]{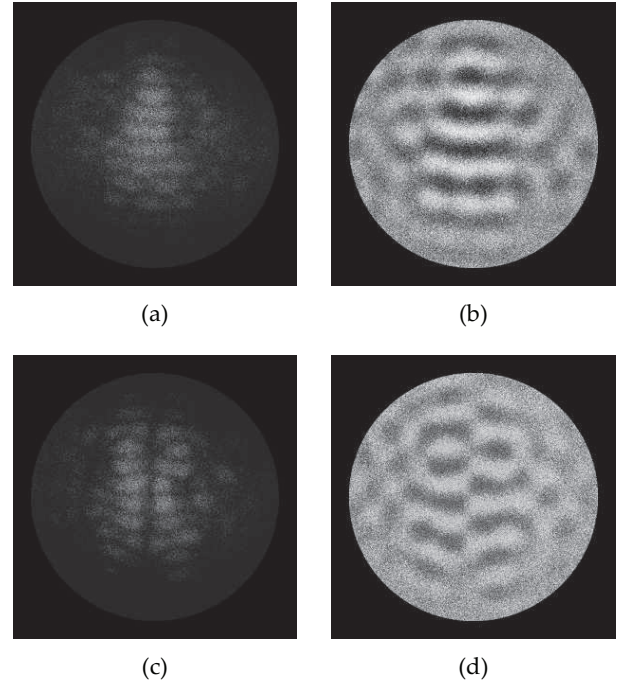}
\caption{Out-of-plane vibration maps of a silicon wafer excited at 142 kHz (a,b), 151 kHz (c,d). Amplitude maps (a,c), in arbitrary units ; black: nodes, white: antinodes. Phase maps (b,d). Black : 0 rad, white : $\pi$ rad.}
\label{fig_wafer}
\end{figure}
Amplitude and phase maps of the vibration modes of a silicium wafer excited at 142 kHz (a,b) and 151 kHz (c,d), are reported in Fig.\ref{fig_wafer}. Each couple of maps is calculated according to eq.\ref{eq_Q}, from the recording of 16 consecutive interferograms ($m=4$ and $n=4$). Phase maps (b,d) exhibit phase oppositions between adjacent bellies (antinodes), in agreement with the modal distribution of the amplitude maps (a,c).\\

\begin{figure}[t]
\centering
\includegraphics[width = 8 cm]{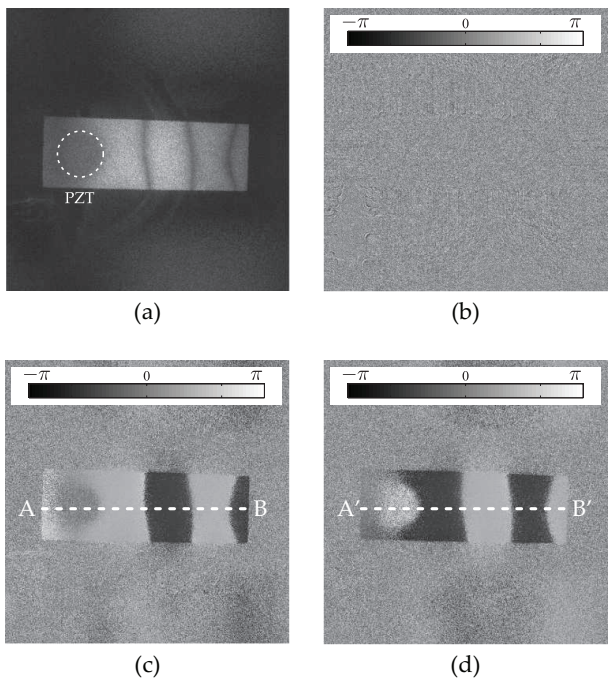}
\caption{Amplitude (a) (black: nodes, white: antinodes) and phase maps (b-d) of a vibrating sheet of paper with no phase shift (c), and $\pi$ phase shift between the excitation and the strobe illumination (d). The phase image rendered without strobe demodulation does not exhibit the vibration phase retardation (b). Dark and light gray regions within the object in (c) and (d) are in phase opposition, as shown in cuts reported in fig\ref{fig_polar}.}
\label{fig_phase}
\end{figure}
\begin{figure}[t!]
\centering
\includegraphics[width = 8.0 cm]{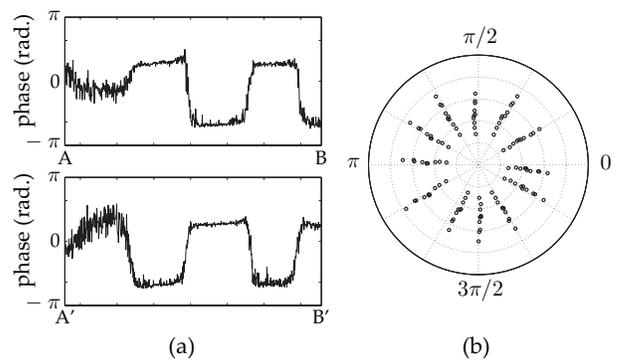}
\caption{(a) Cuts of the vibrating object, along AB and A'B' in Figs. \ref{fig_phase}(c,d). (b) Measured phase of an antinode in fig. \ref{fig_phase} (circles) versus phase detuning between the excitation and the illumination patterns (lines).}
\label{fig_polar}
\end{figure}

The validity of the phase measurement was achieved by estimating the average phase within one antinode of a vibrating sheet of paper, whose lateral dimensions are $9 \times 26 \, \rm mm$, excited at 10 kHz. At this frequency, the steady-state vibrational mode, whose amplitude is reported in Fig.\ref{fig_phase}(a), is composed of rectilinear nodes and bellies oriented vertically \cite{SamsonVerpillat2011}. Black lines are actual antinodes, and not time-averaged Bessel fringes, since the vibration amplitude is much lower than the optical wavelength. Amplitude and phase maps measured for a strobe signal phase-shifted by 0 and $\pi$ radians with respect to the mechanical excitation, respectively, are reported in Fig.\ref{fig_phase}(c) and Fig.\ref{fig_phase}(d). Once again, each couple of maps was calculated according to eq.\ref{eq_Q}, from 16 raw frames ($m=4$ and $n=4$). To avoid any ambiguity of the mechanical phase determination, bursts of camera frames and strobe LO signal patterns $H_{\rm AM} (t)$ were triggered by a clock signal, derived from frequency-division of the mechanical excitation signal. Reconstructed mechanical phases in (c) and (d) exhibit inverted contrast from-antinode-to-antinode, which corresponds to phase opposition. These phase oppositions are also visible in the cuts reported in Fig. \ref{fig_polar}(a). Twelve phase shifts between the strobe illumination and the mechanical excitation signal, linearly-spaced between 0 and $2\pi$ radians are set and measured sequentially. Measures are reported in Fig.\ref{fig_polar}(b). They are in agreement with the set of imposed phase shifts.\\

In conclusion, we demonstrated that quantitative phase maps from sinusoidal out-of-plane vibration can be measured by frequency-shifted holography tuned for first modulation side band detection in conjunction with a stroboscopic illumination. In the reported approach, phase maps are derived from sequences of hologram intensity measures. Its robustness and accuracy lies in the continuous drift of the optical pulses with respect to the mechanical excitation signal, throughout image sequence acquisition. With this method, sequential shifting of strobe pulse is circumvented.\\

This work was funded by Fondation Pierre-Gilles de Gennes (FPGG014 grant), Agence Nationale de la Recherche (ANR-09-JCJC-0113, ANR-11-EMMA-046 grants), and R\'egion \^Ile-de-France (C'Nano grant, AIMA grant).


\begin{thebibliography}{0}
\expandafter\ifx\csname natexlab\endcsname\relax\def\natexlab#1{#1}\fi
\expandafter\ifx\csname bibnamefont\endcsname\relax
  \def\bibnamefont#1{#1}\fi
\expandafter\ifx\csname bibfnamefont\endcsname\relax
  \def\bibfnamefont#1{#1}\fi
\expandafter\ifx\csname citenamefont\endcsname\relax
  \def\citenamefont#1{#1}\fi
\expandafter\ifx\csname url\endcsname\relax
  \def\url#1{\texttt{#1}}\fi
\expandafter\ifx\csname urlprefix\endcsname\relax\def\urlprefix{URL }\fi
\providecommand{\bibinfo}[2]{#2}
\providecommand{\eprint}[2][]{\url{#2}}

\end{thebibliography}


\begin{thebibliography}{10}
\newcommand{\enquote}[1]{``#1''}

\bibitem{Powell1965}
R.~L. Powell and K.~A. Stetson, J. Opt. Soc. Am. \textbf{55}, 1593 (1965).

\bibitem{Levitt1976}
J.~Levitt and K.~Stetson, Applied Optics \textbf{15}, 195 (1976).

\bibitem{PicartLeval2003}
P.~Picart, J.~Leval, D.~Mounier, and S.~Gougeon, Opt. Lett. \textbf{28}, 1900
  (2003).

\bibitem{Pedrini2006}
G.~Pedrini, W.~Osten, and M.~E. Gusev, Appl. Opt. \textbf{45}, 3456 (2006).

\bibitem{Aleksoff1971}
C.~C. Aleksoff, Applied Optics \textbf{10}, 1329 (1971).

\bibitem{JoudLaloe2009}
F.~Joud, F.~Lalo\"{e}, M.~Atlan, J.~Hare, and M.~Gross, Opt. Express
  \textbf{17}, 2774 (2009).

\bibitem{Lokberg1976}
O.~J. L{\o}kberg and K.~H{\o}gmoen, Appl. Opt. \textbf{15}, 2701 (1976).

\bibitem{Petitgrand2001}
S.~Petitgrand, R.~Yahiaoui, K.~Danaie, A.~Bosseboeuf, and J.~Gilles, Optics and
  lasers in engineering \textbf{36}, 77 (2001).

\bibitem{LevalPicart2005}
J.~Leval, P.~Picart, J.~P. Boileau, and J.~C. Pascal, Appl. Opt. \textbf{44},
  5763 (2005).

\bibitem{Hernandez2011}
M.~Hernandez-Montes, F.~Mendoza~Santoyo, C.~Perez~Lopez,
  S.~Mu{\~n}oz~Sol{\'\i}s, and J.~Esquivel, Optics and Lasers in Engineering
  (2011).

\bibitem{AtlanGross2007}
M.~Atlan, M.~Gross, and E.~Absil, Optics Letters \textbf{32}, 1456 (2007).

\bibitem{HennellySheridan2005}
B.~M. Hennelly and J.~T. Sheridan, J. Opt. Soc. Am. A \textbf{22}, 928 (2005).

\bibitem{SamsonVerpillat2011}
B.~Samson, F.~Verpillat, M.~Gross, and M.~Atlan, Opt. Lett. \textbf{36}, 1449
  (2011).

\end{thebibliography}

\end{document}